\documentclass[prd,twocolumn,a4paper,superscriptaddress,floatfix]{revtex4}
\usepackage{bbm}
\usepackage[all]{xy}
\usepackage{amsmath}
\usepackage{amssymb}
\usepackage[pdftex]{graphicx}
\usepackage{epstopdf}

\def\be{\begin{equation}}
\def\ee{\end{equation}}
\newcommand{\bes}{\begin{subequations}}
\newcommand{\ees}{\end{subequations}}

\def\ben{\begin{eqnarray}}
\def\een{\end{eqnarray}}
\def\ba{\begin{array}}
\def\ea{\end{array}}

\begin{document}
\newcommand{\half}{{\textstyle\frac{1}{2}}}
\allowdisplaybreaks[3]
\def\a{\alpha}
\def\b{\beta}
\def\g{\gamma}\def\G{\Gamma}
\def\d{\delta}\def\D{\Delta}
\def\ep{\epsilon}
\def\et{\eta}
\def\z{\zeta}
\def\t{\theta}\def\T{\Theta}
\def\l{\lambda}\def\L{\Lambda}
\def\m{\mu}
\def\f{\phi}\def\F{\Phi}
\def\n{\nu}
\def\p{\psi}\def\P{\Psi}
\def\r{\rho}
\def\s{\sigma}\def\S{\Sigma}
\def\ta{\tau}
\def\x{\chi}
\def\o{\omega}\def\O{\Omega}
\def\k{\kappa}
\def\pa {\partial}
\def\ov{\over}
\def\br{\\}
\def\ud{\underline}

\newcommand\lsim{\mathrel{\rlap{\lower4pt\hbox{\hskip1pt$\sim$}}
    \raise1pt\hbox{$<$}}}
\newcommand\gsim{\mathrel{\rlap{\lower4pt\hbox{\hskip1pt$\sim$}}
    \raise1pt\hbox{$>$}}}
\newcommand\esim{\mathrel{\rlap{\raise2pt\hbox{\hskip0pt$\sim$}}
    \lower1pt\hbox{$-$}}}

\title{Impact of the matter density uncertainty on the dark energy reconstruction}

\author{P.P. Avelino}
\email[Electronic address: ]{ppavelin@fc.up.pt}
\affiliation{Centro de F\'{\i}sica do Porto, Rua do Campo Alegre 687, 4169-007 Porto, Portugal}
\affiliation{Departamento de F\'{\i}sica da Faculdade de Ci\^encias
da Universidade do Porto, Rua do Campo Alegre 687, 4169-007 Porto, Portugal}

\begin{abstract}

In this paper we study the impact of the fractional matter density uncertainty in 
the reconstruction of the equation of state of dark energy. We consider both 
standard reconstruction methods, based on the dynamical effect that dark energy has on the expansion of 
the Universe, as well as non-standard methods, in which the evolution of the dark energy equation 
of state with redshift is inferred through the variation of fundamental couplings such as the fine structure 
constant, $\alpha$, or the proton-to-electron mass ratio, $\mu$. We show that  the negative impact of the 
matter density uncertainty in the dark energy reconstruction using varying couplings may be very small compared 
to standard reconstruction methods. We also briefly discuss other fundamental questions which need to be 
answered before varying couplings can be successfully used to probe the nature of the dark energy.

\end{abstract} 
\pacs{98.80.Cq}
\maketitle

\section{Introduction}

Multiple observations in the last decade have provided very strong evidence for a recent acceleration of the universe 
\cite{Perlmutter:1998np, Riess:1998cb,Riess:2004nr,Riess:2006fw,Percival:2007yw,Komatsu:2008hk}. 
In the context of Einstein General Relativity such an acceleration can only be explained by an exotic dark energy component violating the strong energy condition. From a purely phenomenological point of view the cosmological 
constant appears to be the simplest dark energy candidate. Still, a dynamical scalar field is expected to be a more plausible explanation 
particularly in face of the very large discrepancy between the observationally inferred vacuum energy density and theoretical expectations. 

If the dark energy is dynamical a fundamental question immediately arises regarding the characterization of the 
evolution of its properties with redshift and in particular of its equation of state. Standard methods to reconstruct the dark energy equation of state as a function of the redshift rely on the dynamical effect that dark energy has on the expansion of the 
universe \cite{Albrecht:2009ct}. However, dark energy is dynamically relevant mainly at recent times which makes the task of accurately determining the evolution of its equation of state at $z \gsim 1$ an almost impossible one, at least using standard methods.

On the other hand, cosmological variations of fundamental couplings can be probed over a wide redshift range. At high redshift cosmic microwave background temperature and polarization anisotropies  \cite{Avelino:2000ea,Avelino:2001nr,Martins:2003pe,Rocha:2003gc, Stefanescu:2007aa,Nakashima:2008cb,Landau:2008re,Scoccola:2008jw} and light element abundances \cite{Bergstrom:1999wm,Avelino:2001nr,Nollett:2002da,Landau:2008re} constrain the value of $\alpha$ at $z \sim 10^{10}$ and $z \sim 10^3$ to be within a few percent of its present day value. Positive results suggesting a cosmological variation of the fine-structure constant, $\alpha$, and the proton-to-electron mass ratio, $\mu$, at about the $10^{-5}$ have been reported  in the redshift range $z=1-4$ \cite{Webb:1998cq,Murphy:2006vs,Ivanchik:2005ws,Reinhold:2006zn}. Unfortunately, other analysis have found no evidence for such variations \cite{Chand:2004ct,Molaro:2007kp,Wendt:2008fx}. At even lower redshifts laboratory experiments and the Oklo natural nuclear reactor provide very stringent limits on the time-variation of $\alpha$ and $\mu$ \cite{Peik:2006xy,Blatt:2008su,Rosenband,Gould:2007au}.

 If the dark energy is described by a quintessence field, $\phi$, non-minimally coupled to the electromagnetic field \cite{Carroll:1998zi,Olive:2001vz,Chiba:2001er,Wetterich:2002ic,Parkinson:2003kf,Anchordoqui:2003ij,Copeland:2003cv,Nunes:2003ff,Avelino:2004hu,Doran:2004ek,Marra:2005yt,Avelino:2005pw,Avelino:2008dc} then the dynamics of $\alpha$ is coupled to the evolution of $\phi$. 
It was shown \cite{Avelino:2006gc} that, under certain assumptions, varying couplings may be used to probe the nature of dark energy over a larger redshift range than that spanned by standard methods (such as 
supernovae \cite{Perlmutter:1998np,Riess:1998cb,Riess:2004nr,Riess:2006fw} or weak lensing \cite{Huterer:2001yu,Munshi:2006fn,Hoekstra:2008db}). However, a perfect knowledge of $\Omega_{m0}$ was assumed which can not be realized in practice. In the present paper we eliminate this shortcoming by studying the impact that the matter density uncertainty has on the dark 
energy reconstruction using varying couplings.

This paper is organized as follows. In Section II we start by quantifying the impact of the fractional matter 
density uncertainty in the determination of the evolution of the dark energy equation of state with redshift using 
standard reconstruction methods. In Section III we perform a quantitative analysis of the same problem now assuming that 
varying couplings 
are used to probe the nature of dark energy. We consider a broad class of models for the evolution of $\alpha$ and $\mu$ 
where the gauge kinetic function is a linear function of a quintessence-type real scalar field described by a Lagrangian with a standard kinetic term and a scalar field potential, $V(\phi)$. We conclude in Section IV with a brief summary of our results and a discussion of future prospects. 

Throughout this paper we shall use units with $\hbar=c=8 \pi G=1$ and a metric signature $(+,-,-,-)$.

\section{Standard dark energy reconstruction}

Consider a flat homogeneous and isotropic Friedmann-Robertson-Walker universe whose dynamics 
is described by
\begin{eqnarray}
\label{fried1}
H^2 = \frac{\rho}{3}\,,\\
\label{fried2}
HH' = -\frac{1}{2} \left(\rho+p\right) \,,
\end{eqnarray}
where $\rho$ is the total density, $p$ is the total pressure, $a$ is the scale factor, a prime represents a derivative with respect to $\ln a$, $H={\dot a}/a$ is the Hubble parameter and a dot represents a derivative with respect to physical time, $t$. 
Eqs.  (\ref{fried1}) and (\ref{fried2}) can also be combined to give
\begin{equation}
\label{fried3}
w\equiv \frac{p}{\rho}=-\frac23 \frac{H'}{H} -1\,.
\end{equation}
Eq. (\ref{fried3}) tell us that a precise knowledge of the universe dynamics (and therefore of the evolution of $H$) is 
all that is required in order to determine the evolution of $w$. 
If the universe contains only minimally coupled fluids energy-momentum conservation implies that the equation
\begin{equation}
\label{energyc}
\rho' +3(\rho+p)=0 \,,
\end{equation}
is satisfied not only by the mixture but also by each individual fluid.
In this paper we will make the simplifying assumption that the universe is constituted solely by matter and dark energy with 
energy densities $\rho_m$ and $\rho_\phi$, thus neglecting the residual (at recent times) radiation component.
Consequently, $\rho=\rho_m+\rho_\phi$ and $p=p_\phi$ (taking $p_m=0$) so that
\begin{equation}
w=\frac{w_\phi}{1+\Omega_m/\Omega_\phi}=w_\phi (1-\Omega_m)\,,
\end{equation}
where $\Omega_m=\rho_m/\rho$, $\Omega_\phi=\rho_\phi/\rho$ and $\Omega_m+\Omega_\phi=1$ since we are 
considering a flat universe.

Standard methods to reconstruct the dark energy equation of state rely on the dynamical effect that dark energy  has on the 
expansion of the universe. They are faced with two main limitations. The first is related to the fact that the dynamics of the universe is not known with infinite precision. Hence, at each redshift, $z$, the value of $w$ will have an uncertainty $\Delta w$ that will translate into a much larger uncertainty in $\Delta w_\phi$
\begin{equation}
\Delta w_\phi = \Delta w \left(1+\frac{\Omega_m}{\Omega_\phi} \right)= \frac{\Delta w}{1-\Omega_m}\,.
\end{equation}
For the moment we are assuming no uncertainty in $\Omega_m$ (i.e. $\Delta \Omega_m=0$). 

The ratio 
$\Omega_m/\Omega_\phi$ grows rapidly with redshift at recent times and consequently we expect that 
$\Delta w_\phi \gg \Delta w$ for $z \gsim 1$ (note that $1+z =1/a$).
If we assume that the dark energy is a cosmological constant then
\begin{equation}
1+\frac{\Omega_m}{\Omega_\phi}=1+\frac{\Omega_m}{\Omega_\Lambda}=
1+\frac{\Omega_{m0} (1+z)^3}{\Omega_{\Lambda 0}} \sim 1+ 0.37  (1+z)^3\,,
\end{equation}
which becomes a very large factor at high redshift (here we took $\Omega_{m0}=0.27$ and $\Omega_{\Lambda 0}=0.73$ as  favored by the five-year WMAP results \cite{Komatsu:2008hk}). The  subscript `0' indicates that the variables are 
to be evaluated at the present epoch.

\begin{figure}[t!]
\includegraphics[width=7.5cm]{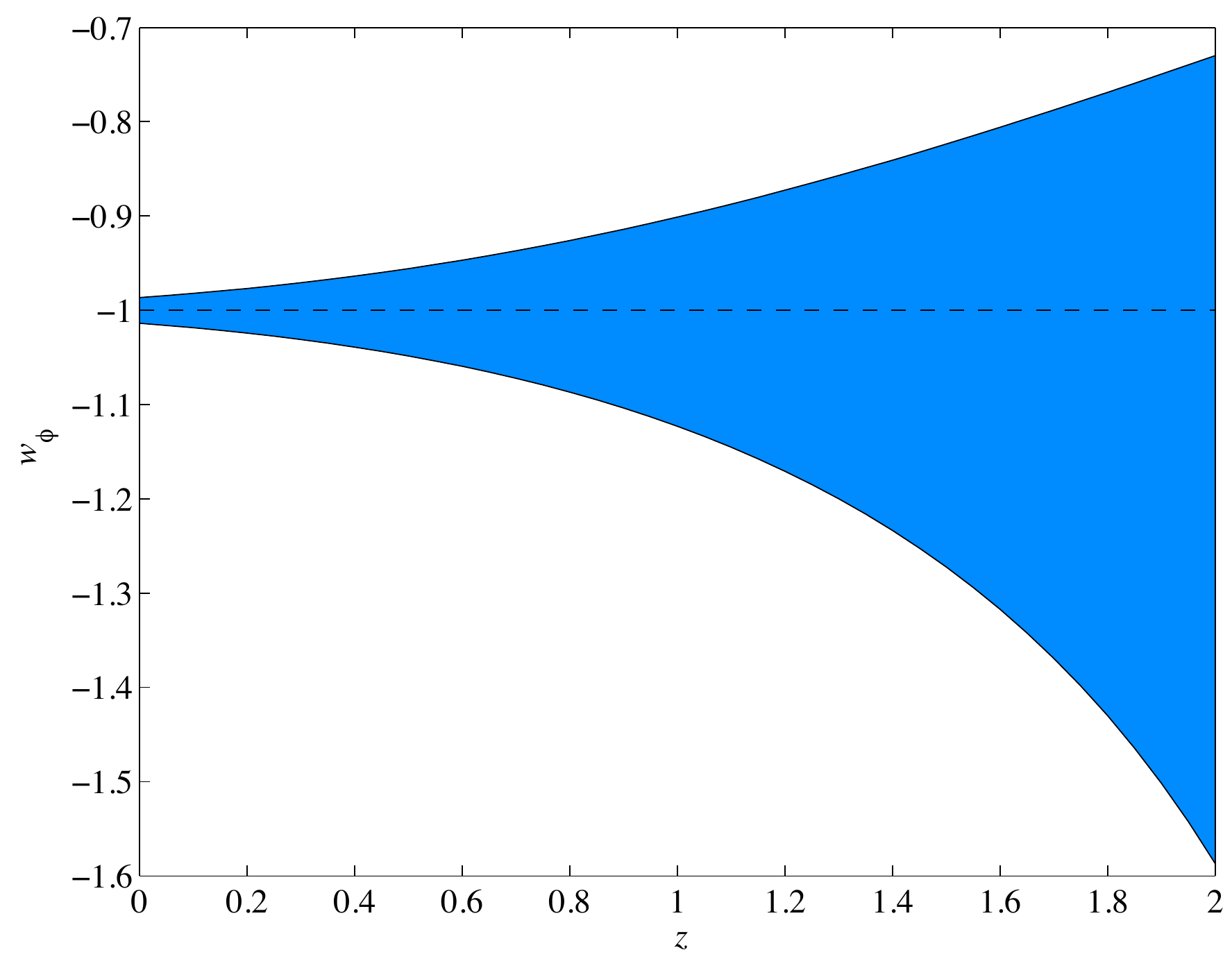}
\caption{\small{Evolution with redshift, $z$, of the uncertainty in the value of $w_\phi$ assuming a background $\Lambda {\rm CDM}$ model with $\Omega_{m0}=0.27$, $\Omega_{\phi 0}=\Omega_{\Lambda 0}=0.73$ and an observational uncertainty $\Delta \Omega_{m 0}=0.01$ in the fractional matter density. The dashed line represents the correct value of the dark energy equation of state parameter (equal to $-1$ independently of $z$, in this particular case).}}
\end{figure}

A second source of uncertainty in $w_\phi$ is due to the fact that we do not know $\Omega_{m 0}$ with infinite precision \cite{Wasserman:2002gb,Kunz:2007nn,Sahni:2008xx}. This uncertainty would be present even if we had a perfect knowledge of the evolution of $w$ (in which case $\Delta w$ would be zero). Although this ideal situation cannot be achieved in practice, we will consider it for the sake of illustration. Still, a perfect knowledge of the dynamics of $w$ would not 
imply that the evolution of $w_\phi$ could also be determined with arbitrary precision. An uncertainty in the value of 
matter density, $\Delta \rho_m$, will be reflected in a corresponding uncertainty in the value of $\rho_\phi$, since we are unable to determine, with absolute certainty, whether or not a fraction of the matter density is really dark matter or dark energy. The corresponding uncertainty in the dark energy equation of state parameter is equal to
\begin{eqnarray}
\Delta w_\phi&=&\frac12\left| \frac{p_\phi}{\rho_\phi-\Delta \rho_m}- \frac{p_\phi}{\rho_\phi+\Delta \rho_m}\right| \sim
\frac{|w_\phi| \Delta \Omega_m}{1-\Omega_m}\nonumber \\
&=& \frac{|w| \Delta \Omega_m}{(1-\Omega_m)^2} \,.
\end{eqnarray}
where $\Delta \Omega_m=\Delta \rho_m/\rho>0$ and the approximation is valid only if $\Delta w_\phi \ll |w_\phi|$. If the dark energy is a cosmological constant then we have
\begin{equation}
\Omega_m = \frac{\Omega_{m0}}{\Omega_{m0} + \Omega_{\Lambda 0}(1+z)^{-3}},.
\end{equation}
with $\Omega_\phi=\Omega_\Lambda=1-\Omega_m$. On the other hand we assume that
\begin{equation}
\frac{\Delta \Omega_m}{\Omega_m} = \frac{\Delta \Omega_{m0}}{\Omega_{m0}}\,.
\end{equation}

In Fig.~1 we plot the evolution with redshift, $z$, of the uncertainty in the dark energy equation of state parameter, $[w_\phi-\Delta w_\phi,w_\phi+\Delta w_\phi]$, assuming again that 
$\Omega_{m0}=0.27$, $\Omega_{\phi 0}=\Omega_{\Lambda 0}=0.73$ and an observational uncertainty $\Delta \Omega_{m0}=0.01$ in the fractional matter density. The dashed line represents the correct value of the dark energy equation of state parameter (equal to $-1$ independently of $z$, in this particular case). The figure clearly shows a small uncertainty in the value of $\Omega_{m0}$ at $z=0$ originates a very large uncertainty in the evolution of the dark energy equation state, specially at high redshift.

\section{Dark energy reconstruction using varying couplings}

In this section we shall consider a class of models models where a 
quintessence field is non-minimally coupled to the electromagnetic field. 
These models are defined by the action
\begin{equation}\label{eq:L}
S=\int d^4x \, \sqrt{-g} \mathcal \, {\cal L} \, ,
\end{equation}
where
\begin{equation} 
{\cal L} = {\cal L}_\phi + {\cal L}_{\phi F} + {\cal L}_{\rm other}\, , 
\end{equation} 
${\cal L}_\phi=X-V(\phi)$,
\begin{equation}\label{eq:kinetic_scalar1}
X=\frac{1}{2}\nabla^\mu \phi \nabla_\mu \phi \,,
\end{equation}
$V(\phi)$ is the scalar field potential,
\begin{equation} 
{\cal L}_{\phi F}= -\frac{1}{4}  B_F (\phi) F_{\mu \nu} F^{\mu \nu}\, , 
\end{equation}
$B_F(\phi)$ is the gauge kinetic function, $F_{\mu \nu}$ are the components of 
the electromagnetic field tensor and ${\cal L}_{\rm other}$ is the Lagrangian density of the other fields. The fine-structure constant is given by 
\be 
\alpha(\phi)=\frac{\alpha_0}{B_F(\phi)} 
\label{gkfalpha} 
\ee 
and, at the present day, one has $B_F(\phi_0)=1$. 

We will also assume that the gauge kinetic function is a linear 
function of $\phi$ so that one has 
\be 
\frac{\delta \alpha}{\alpha}=\zeta \delta \phi\,,
\label{lin} 
\ee 
where $\delta \alpha =\alpha_0-\alpha$, $\delta \phi =\phi_0-\phi$ and $\zeta$ 
is a constant. 

The evolution of $\phi$ induced solely by its coupling to electromagnetically interacting matter is so small (given Weak Equivalence Principle constraints \cite{Olive:2001vz,Schlamminger:2007ht,Dent:2008vd}) that the resulting time variation of $\alpha$ can be neglected. Hence, throughout this paper we shall assume that the 
dynamics of $\phi$ is fully driven by the scalar field potential, $V(\phi)$ 
(and damped by the expansion). Throughout this paper we shall also neglect the spatial variations of $\phi$ 
which was shown to be a good approximation in this context \cite{Shaw:2005gt,Avelino:2005pw,Avelino:2008cu}. 

The relation between the variations of $\alpha$ and $\mu$ is model dependent 
but, in general, we expect that
\begin{equation}
\frac{\dot \mu}{\mu}=R \frac{\dot \alpha}{\alpha}
\end{equation}
where  $R$ is a model dependent constant (see \cite{Campbell:1994bf,Calmet:2001nu,Langacker:2001td,Calmet:2002ja,Olive:2002tz,Dine:2002ir,Dent:2008us} for detailed discussion of specific models).

Taking into account that $\rho=\rho_m+\rho_\phi$, $\rho_\phi={\dot \phi}^2/2+V(\phi)$, $p_\phi={\dot \phi}^2/2-V(\phi)$, 
$p_m=0$ and ${\dot \phi} = \phi' H$, Eqs. (\ref{fried1}) and  (\ref{energyc}) can be written as
\begin{eqnarray}
\label{H2}
H^2 = \frac{1}{3} \left(\rho_m + \rho_\phi\right) \,,\\
\label{ec}
\rho_\phi' =-3H^2 {\phi'}^2\,,
\end{eqnarray}
or alternatively
\begin{eqnarray}
\label{H21}
H^2 = H_0^2 \Omega_{m0} \left(\sigma + a^{-3}\right)\,,\\
\label{ec1}
\frac{\sigma'}{\sigma_0'} =\left(\frac{H}{H_0}\right)^2 \left(\frac{\phi'}{\phi_0'}\right)^2\,,
\end{eqnarray}
where $\sigma=\rho_\phi/(\rho_0 \Omega_{m0})$ so that 
\begin{equation}
\label{sig0}
\sigma_0=\Omega_{\phi 0}/\Omega_{m0}\,.
\end{equation}
Substituting Eq. (\ref{H21}) into Eq. (\ref{ec1}) one obtains \cite{Nunes:2003ff}
\begin{eqnarray}
\label{sigma}
\sigma' = \frac{\sigma_0'}{\sigma_0+1}\ \left(\frac{\phi'}{\phi_0'}\right)^{2} \left(\sigma + a^{-3}\right)
\,.
\end{eqnarray}
Note that the 
linearity assumption given by Eq. (\ref{lin}) implies that $\phi'/\phi_0'=\alpha'/\alpha_0'$.
The dark energy equation of state parameter is given by
\begin{equation}
\label{wphi}
w_\phi = -1 +  \frac{\rho_\phi+p_\phi}{\rho_\phi}=-1-\frac13 \frac{\rho_\phi' }{\rho_\phi}=-1-\frac13 \frac{\sigma' }{\sigma}\,,
\end{equation}
and consequently 
\begin{equation}
\label{sig0d}
\sigma_0'=-3\sigma_0(w_{\phi 0}+1)=-3\sigma_0\left(\frac{w_0}{1-\Omega_{m0}}+1\right)\,. 
\end{equation}
Eq. (\ref{wphi}) can also be written as
\begin{equation}
\label{wphi1}
w_\phi = -1 +  \frac{{\phi'}^2H^2}{\rho_\phi}=-1+ \frac{{\phi'}^2}{3\Omega_\phi}\,.
\end{equation}
which implies that
\begin{equation}
\label{fphi}
f_{\phi} \equiv \frac{w_\phi+1}{w_{\phi 0}+1} = \left(\frac{\phi'}{\phi_0'}\right)^2 \frac{\Omega_{\phi 0}}{\Omega_\phi}=\left(\frac{\alpha'}{\alpha_0'}\right)^2 \frac{\Omega_{\phi 0}}{\Omega_\phi}\,,
\end{equation}
where
\begin{equation}
\label{omegap}
\Omega_\phi=\frac{\rho_\phi}{\rho}=\frac{\sigma H_0^2 \Omega_{m0}}{H^2}=\frac{\sigma}{\sigma+a^{-3}}\,.
\end{equation}

\begin{figure}[t!]
\includegraphics[width=7.0cm]{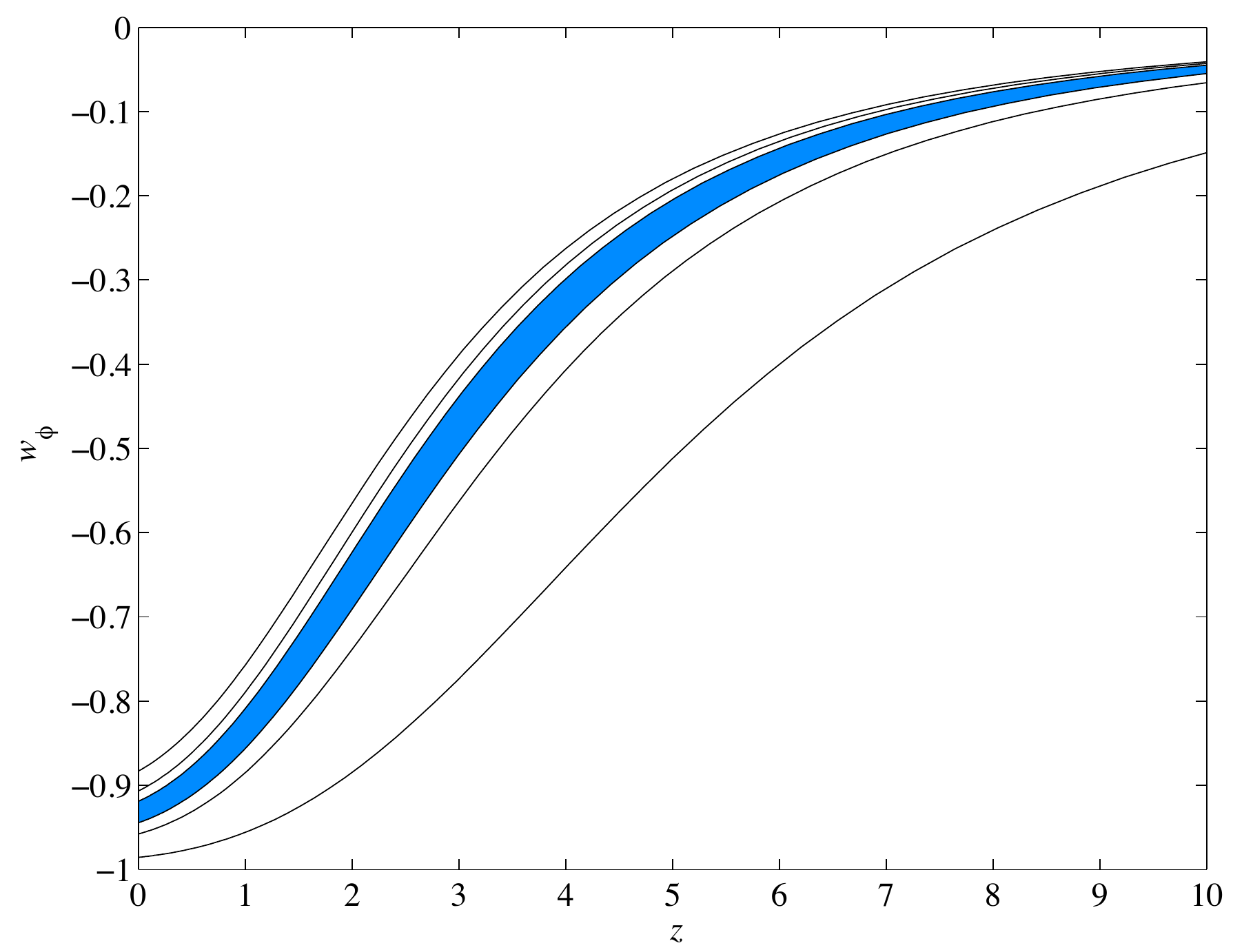}
\\
\vspace{0.3cm}
\includegraphics[width=7.0cm]{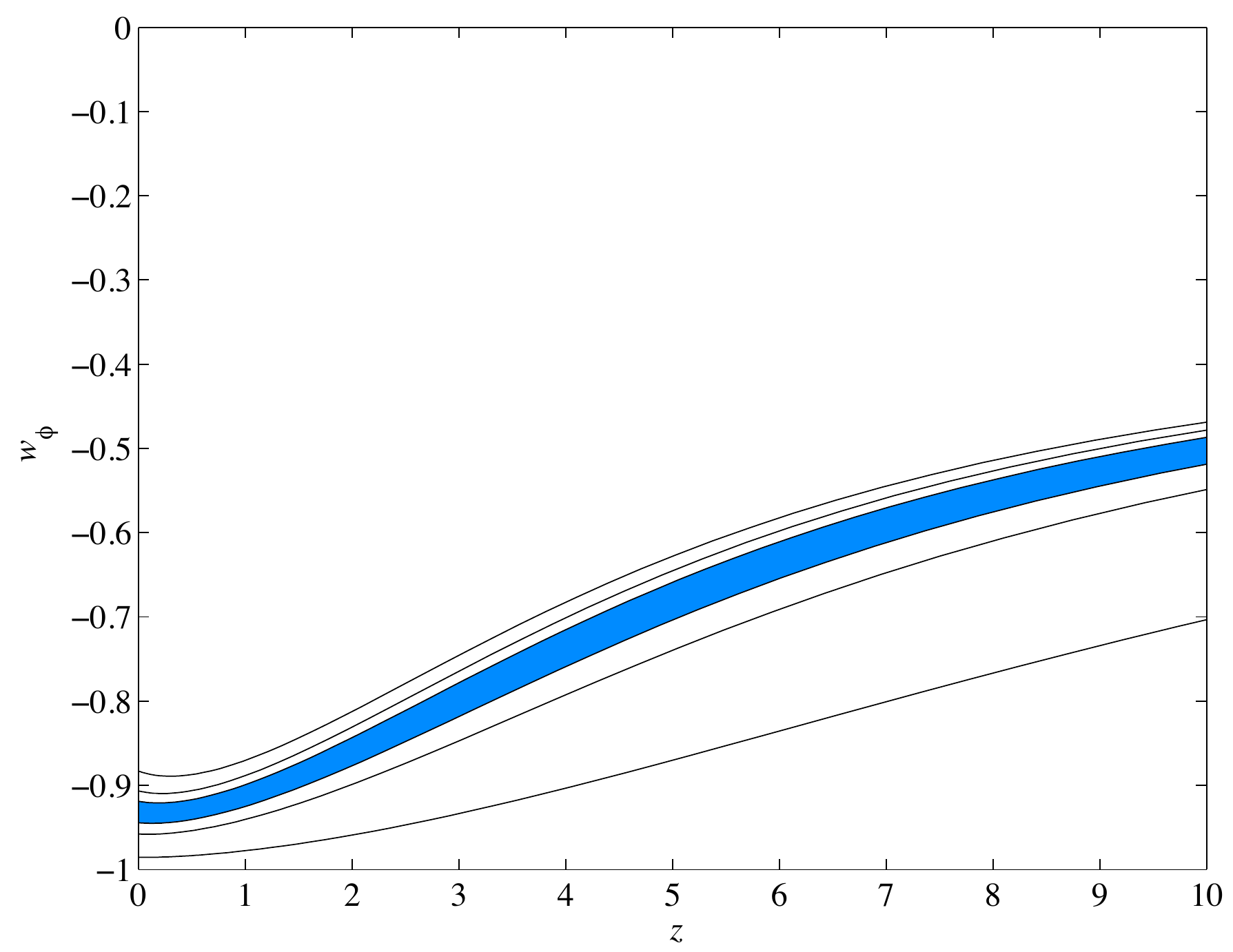}
\caption{\small{The evolution with redshift, $z$, of the equation of state parameter, $w_\phi$, determined using Eqs. (\ref{fphi}) and (\ref{omegap}) and taking $w_0=-0.68$, $\Omega_{m0}=0.27+\delta \Omega_{m0}$ with $\delta \Omega_{m0}=-0.04,-0.02,-0.01,0.01,0.02,0.04$ ($\Omega_{\phi 0}=1-\Omega_{m0}$). The blue stripe represents the interval 
$[w_\phi - \Delta w_\phi,w_\phi + \Delta w_\phi]$ with the lower and upper limits of the interval calculated using 
$\Omega_{m0}=0.26$ and $\Omega_{m0}=0.28$ respectively. Here we consider a model with constant $\alpha'$ (upper panel) 
and another with $\alpha' \propto a^{1/2}$  (lower panel).}}
\end{figure}

In this paper our aim is to isolate the effect of the uncertainty in $\Omega_{m0}$ on our ability to reconstruct the evolution with 
redshift of the dark energy equation of state and consequently we shall take the ideal situation in which the evolution of 
$\alpha$ with redshift is known with infinite precision (note that forecasts taking into account expected measurement 
uncertainties for the next generation of spectrographs have already been presented elsewhere \cite{Avelino:2006gc}).
Hence, here we assume that the uncertainty in $f_\phi (z)$ is completely due to the uncertainty in $\Omega_\phi(z)$ (or equivalently in $\sigma(z)$).

In Fig.~2 we plot the evolution with redshift, $z$, of the equation of state parameter, $w_\phi$, determined using Eqs. (\ref{fphi}) and (\ref{omegap}) and taking $w_0=-0.68$, $\Omega_{m0}=0.27+\delta \Omega_{m0}$ with $\delta \Omega_{m0}=-0.04,-0.02,-0.01,0.01,0.02,0.04$ ($\Omega_{\phi 0}=1-\Omega_{m0}$). The blue stripe represents the interval 
$[w_\phi - \Delta w_\phi,w_\phi + \Delta w_\phi]$ with the lower and upper limits of the interval calculated using 
$\Omega_{m0}=0.26$ and $\Omega_{m0}=0.28$ respectively. 
For the sake of illustration we consider a model with constant $\alpha'$ (upper panel) and another with $\alpha' \propto a^{1/2}$  (lower panel). However, we have verified that our main conclusions will not depend on these particular choices for the evolution of $\alpha$.

We clearly see in Fig.~2 that the error bars on the dark energy equation of state reconstruction (blue stripe) due to the uncertainty in the value of $\Omega_{m0}$ do not tend increase with redshift. We can see that is indeed the case by considering a particular set of models in which $\sigma$ becomes considerably greater than $\sigma_0$ at large $z$. When $z \gsim 1$ we have $\sigma a^3 \lsim 1$ and consequently
\begin{eqnarray}
\label{omegap1}
 \frac{\Omega_\phi}{\Omega_{\phi 0}} &\sim& \frac{\sigma a^3}{\Omega_{\phi 0}}\,,\\
 \label{sigmap1}
 \frac{\sigma'}{\Omega_{\phi 0}} &\sim&  \left(\frac{\alpha'}{\alpha_0'}\right)^{2} \xi_0 a^{-3}
\,,
\end{eqnarray}
where
\begin{equation}
\label{xi}
\xi_0=\frac{\sigma_0'}{\Omega_{\phi 0}(\sigma_0+1)}=-3(1+w_{\phi 0})\,.
\end{equation}
The uncertainty in the value of $\xi_0$ is given by
\begin{equation}
\label{dxi}
(1+w_{\phi 0})\frac{\Delta \xi_0}{|\xi_0|} = \Delta w_{\phi 0}\,.
\end{equation}
If the value of $\sigma$ at high redshift is mainly determined by its evolution at $z \gsim 1$ then Eqs. (\ref{fphi}), (\ref{omegap1})  and (\ref{sigmap1}) imply that 
$f_\phi$ is roughly proportional to $\xi_0^{-1}$ and consequently $d f_\phi/f_\phi \sim -d \xi_0/\xi_0$ for  $z \gsim 1$. On the other hand, from Eq. (\ref{fphi}) we have 
that $d w_\phi=f_\phi(d w_{\phi 0} + (1+w_{\phi 0}) df_\phi/f_\phi)$ and we may now estimate the high redshift uncertainty 
in the equation of state parameter in these models to be
\begin{equation}
 \Delta w_\phi  \sim f_\phi  \left(  \Delta w_{\phi 0} - (1+w_{\phi 0})\frac{\Delta \xi_0}{|\xi_0|} \right) \sim 0
\,.
\end{equation}
Although this approximation is not valid at low redshift or while $\sigma \sim \sigma_0$ it clearly shows that, contrary to standard reconstruction methods, the uncertainty in the equation of state parameter due to the uncertainty in $\Omega_{m0}$ does not blow up at high redshift (assuming that the lower limit of the interval $[w_{\phi 0} - \Delta w_{\phi 0},w_{\phi 0} + \Delta w_{\phi 0}]$ is significantly above $-1$). This is hardly surprising since if we were able to track the evolution of $\rho_\phi \propto \sigma$ with 
redshift with arbitrary precision (up to a normalization factor) then we would also have a perfect of knowledge of $w_\phi(z)$.

However, if we consider values of $w_{\phi 0}$ close to $-1$ 
then the distance between adjacent lines in Fig.~2 (upper and lower panels) tends to increase with redshift. This is related to the fact that as we approach $w_{\phi 0}=-1$ 
we get closer to the cosmological constant case for which $w_\phi=-1$ independently of $z$. Still, we note that an observation of a variation of $\alpha$ at $z=0$ immediately leads to an upper bound on $w_{\phi 0}$ \cite{Dent:2008vd}. In particular, from Eqs. (\ref{lin}) and  (\ref{wphi1}) one obtains
\begin{equation}
w_{\phi 0}=-1+  \left(\frac{\alpha_0'}{\zeta \alpha_0 {\sqrt {\Omega_{\phi 0}}}}\right)^2\,,
\end{equation}
where $\zeta \le 1.8 \times 10^{-4}$ considering a variation of $\alpha$ alone (i.e. fixing all other gauge couplings). The constrains  
on $\zeta$ are in general much stronger when more realistic models, based on the unifications of gauge couplings, are considered \cite{Dent:2008vd}. Hence, if $\alpha_0' \neq 0$ then $w_{\phi 0}$ cannot be arbitrarily close to $-1$. The relation between experimental limits at $z=0$ and varying coupling constraints higher redshifts is of course model dependent. However, if $\phi$ 
is always rolling down a monotonic potential then low-redshift constrains on the evolution of $\alpha$ may already provide stringent limits on the corresponding variations at high-redshift \cite{Avelino:2008dc}. 

The upper panel of Fig.~2 shows that the error bars due to the uncertainty in the value of $\Omega_{m0}$ can become significantly smaller at large redshift. Note that  Eq. (\ref{energyc}) implies that for a constant $\phi'$ the energy density of the scalar field is expected to track the background density 
at early times since 
\begin{equation}
\rho_\phi'=-3{\phi'}^2 H^2 \propto \rho\,.
\end{equation}
If  $|\phi_0'|$ is not too small then $w_\phi \to 0$ deep in the matter era irrespectively  of the initial conditions. This is the main reason why the uncertainty becomes very small at large redshift in the upper panel of Fig.~2. We have also verified that if 
$\phi' \propto a^{1/2}$ (lower panel) then $w_\phi \to -1/3$ asymptotically at large $z$. However, in this case the 
convergence is slower than in the previous one. One should however bear in mind that future spectroscopic determinations are unlikely to be able to probe the dynamics of dark energy beyond $z=5-6$.

\section{Conclusions}

In this paper we quantified the impact of the fractional matter density uncertainty,  $\Delta \Omega_{m0}$,  in 
the reconstruction of the equation of state of dark energy considering both standard and non-standard 
reconstruction methods. Standard methods are very sensitive to this uncertainty since the ratio $\Omega_m/\Omega_\phi$ 
is expected to become very large at high redshift. On the other hand, we have shown that the negative impact of 
the matter density uncertainty may be much smaller in the case of non-standard reconstruction methods in which the evolution of 
the dark energy equation of state with redshift is inferred through the variation of fundamental couplings (such 
as $\alpha$ or $\mu$). 

There are however a number of other important questions which still need to be answered positively if varying couplings 
are to be successfully used to infer the dynamics of dark energy: I. Do $\alpha$ ($\mu$) evolve with cosmic time 
and are such variations observable ? II. Is dark energy a quintessence scalar field, $\phi$, described by a Lagrangian 
with a standard kinetic term and an effective potential which is a function of $\phi$ alone ? III. Are the variations of 
$\alpha$  ($\mu$) linearly coupled to $\phi$ ? 

Presently, the answer to the first question is still controversial. Although a number of results suggest a cosmological variation of $\alpha$ and $\mu$ in the redshift range $z=1-4$ other analysis have found no evidence for such variations. However, this 
situation is expected to be fully clarified in the next few years with the next generation of high resolution spectrographs \cite{Liske:2008zu}. Unfortunately, questions 2 and 3 may be easier to answer negatively than positively (for example if standard and non-standard reconstruction methods give incompatible results). Also, note that the varying-couplings method is in general affected by the observational uncertainty in the present-day equation of state $w_0$, which is necessary to
calibrate the relation between $\phi'$ and $\alpha'$ \cite{Nunes:2003ff}. Still, given the strong limitations of standard reconstruction methods alternative promising approaches, such as those based on the time-variation of fundamental couplings, deserve to be further investigated.

\begin{acknowledgments}

We thank the Cosmology Group at Centro de F{\' \i}sica do Porto and Centro de Astrof{\' \i}sica da Universidade do Porto 
for useful discussions. This work was funded by FCT (Portugal) through grant CERN/FP/83508/2008.

\end{acknowledgments}


\bibliography{alpha}

\end{document}